\documentclass[12pt]{iopart}
\input{epsf.tex}

\begin{document}
\title[Born Reciprocity \& 1/r]{Born reciprocity and the 1/r potential}

\author{R Delbourgo\dag and D. Lashmar}
\address{\dag\ School of Mathematics and Physics, University of Tasmania,
         Private Bag 37 GPO Hobart, AUSTRALIA 7001}

\ead{Bob.Delbourgo@utas.edu.au,\,\,dlashmar@utas.edu.au}

\begin{abstract}
Many structures in nature are invariant under the transformation pair,
({\bf p},{\bf r}) $\rightarrow (b\,{\bf r},-{\bf p}/b)$, where $b$ is some
scale factor. Born's reciprocity hypothesis affirms that this invariance
extends to the entire Hamiltonian and equations of motion. We investigate
this idea for atomic physics and galactic motion, where one is basically
dealing with a $1/r$ potential and the observations are very accurate, so 
as to determine the scale $b\equiv m\Omega$. We find that an $\Omega\sim 
1.5\times10^{-15}$ s$^{-1}$ has essentially no effect on atomic physics but 
might possibly offer an explanation for galactic rotation, without invoking 
dark matter.
\end{abstract}

\submitto{\JPA}

\maketitle

\section{Born's Reciprocity Principle} 
One cannot help but be struck by the way that numerous structures in
physics look the same under the simultaneous substitution between momentum
$\bf p$ and position $\bf r$,
\begin{equation}
 {\bf p} \rightarrow b\,{\bf r}, \qquad {\bf r} \rightarrow -{\bf p}/b,
\end{equation}
where $b$ is a scale with the dimensions of M/T. This applies to the
classical Poisson brackets $\{r_i,p_j\}=\delta_{ij}$, the quantum
commutator brackets $[r_i,p_j]=i\hbar\delta_{ij}$, and the form of the
Hamiltonian equations (classical or quantum), $\dot{r}_i =\partial H
/\partial p_i,\,\,\dot{p}_i =-\partial H/\partial r_i$ and of
the angular momenta, $L_{ij} = r_ip_j-r_jp_i$. It leads to the
concept of phase-space, Fourier transforms and uncertainty relations.
The conjugacy between space and momentum is extensible to energy and
time, this being required for special relativity. However these
observations do not presume that Hamiltonians are invariant under  
transformation (1).

Born's reciprocity principle \cite{B,G} goes one stage further and assumes
that all physical equations of motion are {\em invariant} and not just
covariant under such conjugacy transformations, so that
\begin{equation}
 H({\bf r},{\bf p}) = H(-{\bf p}/b, b\,{\bf r}). 
\end{equation}
At first sight this seems a patently absurd idea for anything but 
oscillators and even there it seems quite silly because it leads to 
fixed frequency for all vibrations, determined by the value of $b$. 
For these reasons and for its failure in accounting for the observed 
particle masses \cite{B} the principle has naturally fallen into disrepute
and has never been taken seriously by physicists. There are also
some fundamental philosophical objections to the idea, which will be
mentioned later. In spite of these very valid criticisms, we wish to
explore the principle and see if we can determine a non-zero value of 
$b$ (which is probably tiny indeed and tied to cosmic scales). The
incentive/reason why we wish to entertain the chance that eq.(2) may
be valid arises also from vibrations. The point is that any oscillator
contains some measure of anharmonicity; how this is manifested depends
on the physical context, but we can be sure that the linear force and its
associated potential $V$ cannot increase without end. For instance we 
might suppose that in reality, for mass $m$, the true potential is
$V(r) \simeq (m\omega^2r^2/2)\exp(-cr^2)$, where $c$ is a small 
parameter which sets the distance at which anharmonicity kicks in, and
$\omega$ is the natural rotational frequency for displacements that are 
not excessive. If we attempt to make $H$ reciprocity invariant, we 
may arrive at
$$2H=(p^2+b^2r^2)/m+m\omega^2[r^2\exp(-cr^2)+(p^2/b^2)\exp(-cp^2/b^2)].$$
It follows that if $b$ is miniscule on ordinary momentum scales, the
dangerous last term is minute, as is the correction to the kinetic energy,
so the standard picture prevails for small or moderate $r$. This example 
indicates we have no right to be so dismissive of Born's principle.

If we view the reciprocity substitution as the transformation
$$\left( \begin{array}{c} {\bf p}\\b\,{\bf r}\end{array}\right)\rightarrow
  \left( \begin{array}{cc} 0 & 1 \\ -1 & 0 \end{array} \right)
  \left( \begin{array}{c} {\bf p} \\ b\,{\bf r} \end{array} \right),$$
we can think of another reciprocity transformation which also leaves
most of our physical structures intact, namely
$$\left( \begin{array}{c} {\bf p}\\b\,{\bf r}\end{array}\right)\rightarrow
  \left( \begin{array}{cc} 0 & -i\\ -i & 0 \end{array} \right)
  \left( \begin{array}{c} {\bf p}\\b\,{\bf r} \end{array} \right), \quad
  \left( \begin{array}{c} H \\ bt \end{array} \right) \rightarrow
  \left( \begin{array}{cc} i & 0\\ 0 & -i \end{array} \right)
  \left( \begin{array}{c} H \\ bt \end{array} \right),\quad
  m\rightarrow im.$$ 

In this paper we will consider a non-relativistic potential which is
fully established on the large-scale (Newtonian gravity) and on the 
small scale (atomic physics), namely $1/r$; it has its roots in graviton 
and photon exchange and has been thoroughly studied over the 
centuries! We shall explore how Born's principle affects it. The 
immediate question is how to make $V(r)\propto 1/r$ compatible with (1). 
The substitution, $1/r\rightarrow 1/\sqrt{r^2+p^2/b^2}$ can be rejected 
outright as it would enormously enhance the velocity dependence for small 
$b$, in fact ridiculously so. However a more reasonable alternative is
$1/r \rightarrow (1/r + b/p)$, since the last term fades out as 
$b\rightarrow 0$. \footnote{If a Yukawa potential $\exp(-\mu r)/r$ is 
modified to obey reciprocity, the additional term is $b\exp(-\mu p/b)/p$
and it becomes negiglible for small $b$; thus nuclear physics is unlikely 
to be affected.} There may exist other choices for making $V$ compatible
with Born's principle, but they are probably less natural than the
proposal:
\begin{equation}
H({\bf r},{\bf p})=\frac{{\bf p}^2+b^2{\bf r}^2}{2m}-
                    \alpha\left(\frac{1}{r}+\frac{b}{p}\right),
\end{equation}
where $\alpha$ signifies the interaction strength ($Ze^2/4\pi\epsilon_0$ 
for a hydrogenic atom or $GMm$ for gravity). At this stage we will
take the sign of $b$ positive even though it hails from 
$1/\sqrt{{\bf p}^2}$ and we are not entirely sure about the sign of the
root because the underlying dynamics (and associated field) is unclear. 
The choice of sign will become firmer in section 4 but it is fully 
consonant with the second form of reciprocity substitution. In short, 
for what it's worth, (3) will be our object of study \footnote{Such
a Hamiltonian in turn spawns a strange-looking Lagrangian when it is
expressed in terms of position and velocity.}. It has the curious, 
if not dubious, feature that for any finite energy and 
$r\neq 0$ separation the speed can never vanish; the minimum $v$ may be
very small, being determined by the tiny constant $b$, so it connotes
particle restlessness (like zitterbewegung) even at the classical level!
The only possibility for the speed to vanish is when the singularity
at $r=0$ is reached. Since $r$ stands for the relative 
distance between the test body and the centre of influence, it makes 
more sense to reinterpret $b=m\Omega$ where $m$ is the reduced mass 
and regard $\Omega$ as the truly universal constant. Doing so will 
ensure that the modified kinetic energy assumes the same form no matter 
how coordinates are chosen:
$$p_1^2/m_1+p_2^2/m_2+m_1\Omega^2r_1^2 + m_2\Omega^2r_2^2 = 
P^2/M + p^2/m+ M\Omega^2R^2 + m\Omega^2r^2.$$
Here $R$ is the centre of mass location, $M=m_1+m_2$ is the total
mass and $P$ is the total momentum. Because of $P,R$ dependence this does 
not mean that $H$ is translation invariant any more than the usual 
Coulomb Hamiltonian is invariant under boosts; it is only {\em covariant} 
under those transformations. Therefore this represents a philosophical 
problem for relativity (even Galilean) and we shall worry about it later.

The paper is set out as follows. Section 2 discusses the classical
problem and trajectories as $b$ is varied. Not unexpectedly we find 
that the distorted orbits precess around the force centre and we 
determine the rate for small $b$; this is much like general relativistic 
corrections to Newtonian gravity. Section 3 deals with the quantum 
version; we find the change in energy levels to first order in $b$ 
by the variational method and perturbation theory---which happen to 
agree with one another. In this way we set limits on the value 
of $b$ so as not to disturb experimental atomic results, namely
$b \leq 10^{-26}$ kg/s, a rather weak conclusion. More stringent 
limits come by looking at galactic scales, where the $1/v$ term 
can influence rotation rates profoundly. Section 4 contains our 
investigations of (3) for galaxies (possessing a supermassive black 
hole at the centre); there we find that the $1/v$ term can yield a 
velocity which at first increases linearly from the centre and then 
steadies out to a constant value, before rising again due to the 
effect of the harmonic $b^2r$ accelaration. Our conclusions and 
continued worries with Born's principle end the paper in Section 5. 

\section{Classical motion} 
The equations of motion arising from the non-relativistic expression
(3) are
\begin{equation}
\dot{\bf r} = {\bf p}(1/m+b\alpha/p^3),\quad 
\dot{\bf p} = -{\bf r}(b^2/m +\alpha/r^3),
\end{equation}
with $H=E$, the conserved ``energy''. This means there is a cubic 
relation between momentum and speed, force and displacement. For
positive $b$ the speed/momentum can never vanish, and if $\alpha>0$ 
(an attractive interaction) the force cannot disappear either. For
the purpose of the ensuing analysis we shall take $b$ to be a small
and positive quantity; many of the results can be continued to
negative $b$ without endangering the steps and we shall anyway
be expressing the (small) precessions of the orbit to first order
in $b$ where the sign is somewhat irrelevant.

It is quite difficult to solve these equations in general but it helps
to use rotational invariance of (4) and the conserved angular
momentum $\ell =|{\bf r}\times{\bf p}|$ to simplify the Hamiltonian
in the usual way:
\begin{equation}
H =\frac{1}{2m}\left[p_r^2 + \frac{\ell^2}{r^2} + b^2r^2\right] -
\alpha\left[\frac{1}{r} + \frac{b}{\sqrt{p_r^2+\ell^2/r^2}}\right]=E;
 \quad p_r \equiv {\bf p}\cdot \hat{\bf r}
\end{equation}
(Because of reciprocity we could equally well have used 
$r_p\equiv{\bf r}\cdot \hat{\bf p}$ in place of $p_r$ and $p$ instead
of $br$.  However the form (5) is more familiar and this is the
framework we shall adopt.) In consequence the radial equations are
\begin{equation}
\dot{r}=\frac{p_r}{m}\left[1 + \frac{\alpha bm}
 {(p_r^2+\ell^2/r^2)^{3/2}}\right], \quad
\dot{p}_r=\frac{\ell^2}{mr^3}\!-\!\frac{b^2r}{m}\!-\!\frac{\alpha}{r^2}
 \!+\!\frac{\alpha b\ell^2}{(r^2p_r^2+\ell^2)^{3/2}}.
\end{equation}
For determining the trajectory in the orbital plane, remember that 
the rate of change of azimuthal angle is $\dot{\varphi}=\ell/mr^2$,
so the trajectory equation is obtained by integrating
\begin{equation}
\frac{d\varphi}{dr}= \frac{\ell}
  {r^2p_r\left[1 + \alpha bm/(\ell^2/r^2 + p_r^2)^{3/2}\right]},
\end{equation}
in which $p_r$ has to be eliminated in terms of $E$ via (5). This is
a hard problem for general $b$ so we shall first turn to Mathematica
for some elucidation of the motion and orbits. 

By looking at the $E$-contours in phase space ($r-p_r$) one finds
that for $\ell\neq 0$ the trajectories at lower energy are bounded with 
$p_r=\dot{r}=0$ at perigee or apogee. Absolute minima of $E$ arise
when $\partial E/\partial r = 0$ or where
$$ -\frac{\ell^2}{mr^3}+ \frac{b^2r}{m} + \frac{\alpha}{r^2}
 -\frac{\alpha b}{\ell} = 0.$$
This cubic equation in $r$ is readily solved and it has three real 
roots in the region of interest,
\begin{equation}
r_\pm = \frac{1}{2b}\left[\frac{\alpha m}{\ell} \pm 
\sqrt{\frac{\alpha^2m^2}{\ell^2}-4bl}\right],\quad 
r_0=\sqrt{\frac{\ell}{b}},
\end{equation}
with corresponding extremal energy values
\begin{equation}
E_\pm=-\frac{1}{2m}\left(\frac{\alpha^2m^2}{\ell^2}+2\ell b\right),\quad
E_0 = b\ell-2\alpha\sqrt{\frac{b}{\ell}}.
\end{equation}
$E_0$ is an unstable maximum, while $E_\pm$ are equal value stable
minima, and this is best revealed by plotting $E(r,p_r=0)$ against $r$ 
in Figure 1. The phase space portrait is drawn in Figure 2 and the 
corresponding trajectory is depicted in Figure 3 over a timespan of
two seconds. In all diagrams we have assumed unit mass 
and taken exaggerated values, $\ell =1, \alpha = 10, b=1$ in order
to emphasize the misshapen and precessing orbit.

\begin{figure}
\begin{center}
\epsfbox{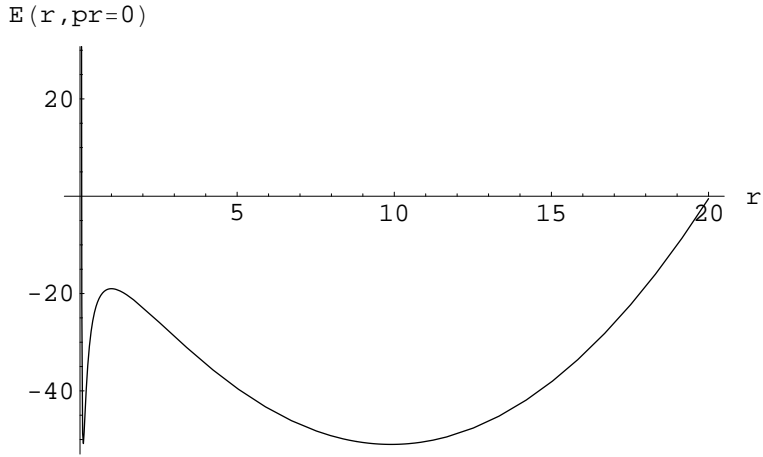}
\end{center}
\caption{\label{Fig1Label}Dependence of energy on radius when
the radial velocity vanishes.}
\end{figure}

\begin{figure}
\begin{center}
\epsfbox{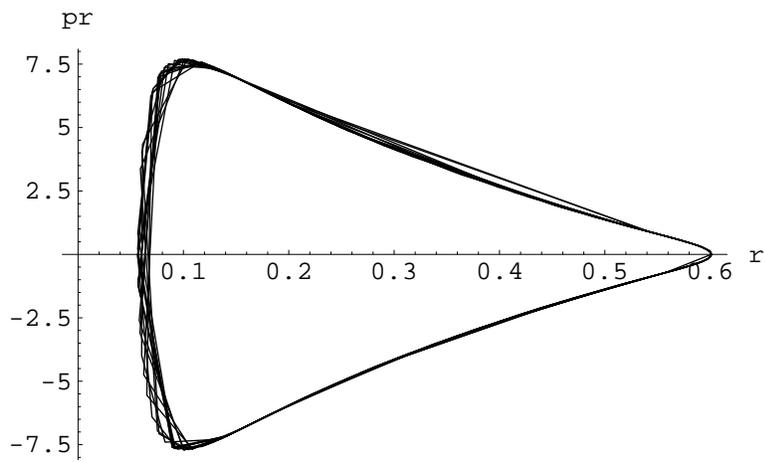}
\end{center}
\caption{\label{Fig2Label}Phase portrait when $E =-21$.}
\end{figure}

\begin{figure}
\begin{center}
\epsfbox{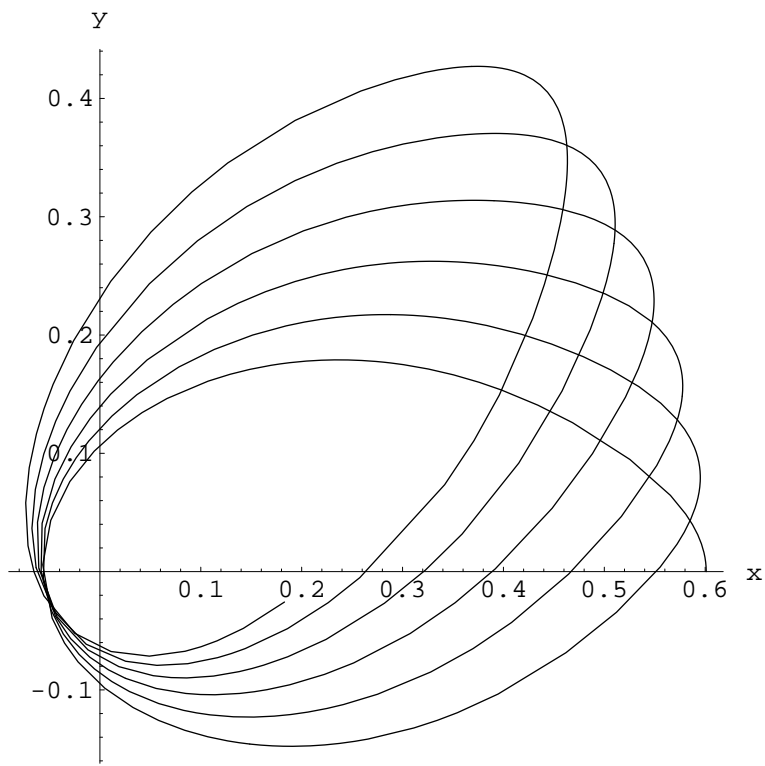}
\end{center}
\caption{\label{Fig3Label}Trajectory when $E=-21$.}
\end{figure}

It is apparent that for $b\neq 0$ orbits become distorted (sometimes
very pronouncedly) from Keplerian ellipses and precession occurs.
It is of interest to work out the precessional rate for model (3) 
and small $b$ when the distortions/changes are tiny too. 
To do so we make the standard change of variable $u = 1/r$ and 
expand (5) and (3) to first order in $b$. Thus
\begin{equation}
\frac{d\varphi}{du} = \frac{\ell}{p_r}\left[1 + \frac{\alpha bm}
  {(\ell^2u^2+p_r^2)^{3/2}}\right]^{-1}\simeq\frac{\ell}{p_r}
 \left[1 - \frac{\alpha bm}{(\ell^2u^2+p_r^2)^{3/2}}\right],
\end{equation}
and 
$$p^3 -2m(E+\alpha u)p - 2m\alpha b + b^2u^{-2}=0; \quad 
 p =\sqrt{p_r^2+\ell^2u^2}.$$
The root we need is $p_r \simeq \sqrt{2m(E+\alpha u)-\ell^2u^2}$,
so expanding around that the trajectory equation (10) simplifies to
\begin{eqnarray}
\frac{d\varphi}{du}&\simeq&\frac{\ell}{\sqrt{2m(E+\alpha u)-\ell^2u^2}}
\left[ 1 - \frac{\alpha bm}{[2m(E+\alpha u)]^{3/2}}\right]\nonumber \\
&=& \frac{1}{\sqrt{(u-u_1)(u_2-u)}}\left[1 - \frac{\alpha bm\ell^3}
 {[(u_1+u_2)u-u_1u_2]^{3/2}}\right] ,
\end{eqnarray}
where $u_1 = 1/r_1, \phi_1=0$ at apogee and $u_2=1/r_2,\phi_2=\pi/2$
at perigee -- the turning points in $r$. The first term on the right of
(11) produces the usual elliptical orbit answer
$$2u = (u_1+u_2) + (u_1-u_2)\cos(2\varphi)$$
and the integration of the second term is connected with the precession.
Over an orbit the additional change in azimuth is 
\begin{eqnarray}
 -\Delta\varphi_b &=& \alpha b m\ell^3\int_{u_1}^{u_2}\frac{du}
 {\sqrt{(u-u_1)(u_2-u)}}\frac{1}{[(u_1+u_2)u-u_1u_2]^{3/2}}\nonumber \\
 &=& \alpha b m\ell^3
 \int_0^\pi \frac{d\phi}{\frac{1}{2}[(u_1^2+u_2^2)+
 (u_1^2-u_2^2)\cos\phi])^{3/2}}\nonumber \\
 &=& \frac{2\alpha bm\ell^3}{u_1^2u_2}E(1-u_1^2/u_2^2) =
     2\alpha bm\ell^3r_1^2r_2 E(1-r_2^2/r_1^2).
\end{eqnarray}
Here $E(k)$ is the complete elliptic integral, which for small argument 
behaves as $E(k) = (\pi/2)(1 - k/4 -\ldots)$ as $k\rightarrow 0$. 
Hence for small eccentricity ($r_1 \simeq r_2 \simeq a$), the change in 
azimuth is
\begin{equation}
\Delta\varphi_b \simeq -2\pi\alpha b m\ell^3a^3,
\end{equation}
to a good approximation, where $a$ is the semimajor axis. In the 
following section we shall set limits on $b$ (for electrons at least)
such that accurate atomic physics experiments are not substantially
disturbed.

\section{Quantum mechanical considerations} 
We will now be dealing with the operator version of (3) for hydrogenic
atoms when $\alpha = Ze^2/4\pi\epsilon_0$.  and
\begin{equation}
H = \frac{P_r^2 +b^2R^2 + L^2/R^2}{2m}-\alpha\left[ \frac{1}{R}
    +\frac{b}{\sqrt{P_r^2+L^2/R^2}}\right].
\end{equation}
It is evident that we are dealing with a tricky problem due to the
last term, connected with $1/|P|$, even for eigenfunctions of angular
momentum $Y_{\ell m}(\theta,\phi)$ so $L^2\rightarrow\ell(\ell+1)\hbar^2$ 
in the Schr\"{o}dinger equation.

Any serious attempt to try to solve (14) needs an interpretation of 
$P_r^{-1}$ when $\ell=0$. As $P_rR(r)\rightarrow(-i\hbar/r)\partial(rR(r))
/\partial r$, when acting on the radial part $R(r)$ of the wave
function, a reasonable definition of the inverse is the indefinite
integral:
\begin{equation}
P_r^{-1}R(r)\rightarrow-\frac{i}{\hbar r}\int_r^\infty dr'\,r'R(r')
\end{equation}
for solutions where $rR(r)\rightarrow 0$ as $r\rightarrow\infty$. One
readily checks that $P_rP_r^{-1}\psi = P_r^{-1}P_r\psi = \psi$ for
normalizable wave functions. It is even true that $P_r^{-1}.\exp(ikr)/r
=(1/\hbar k).\exp(ikr)/r$ for outgoing waves, giving extra credence to
the interpretation (15).

In the event we have not succeeded in obtaining a complete solution 
of $u(r)=r\psi(r)$ of type $P(r)\exp(-\kappa r - br^2/2\hbar)$ to
the equation for the radial wavefunction 
$$-\frac{\hbar^2}{2m}\frac{d^3u}{dr^3}+\frac{d}{dr}\left(
  \frac{b^2r^2u}{2m}-\frac{\alpha u}{r}\right)-\frac{i\alpha bu}{\hbar}
  = E \frac{du}{dr},$$
because of the troublesome last term on the lefthand side. No matter. 
One can still obtain a sensible estimate for the
change in energy levels to $O(b)$ either by using perturbation theory
or making a simple variational calculation. Since the effect is greatest
for the ground state, we will analyse the displacement for the lowest
energy level both ways -- whose answers fortunately agree.

To apply perturbation theory, take the unperturbed wave function in
coordinate and momentum space:
\begin{equation}
\psi(r) = \frac{{\rm e}^{-r/a}}{\sqrt{\pi a^3}},\quad
\phi(p) = \frac{8\sqrt{\pi a^3}}{(p^2a^2/\hbar^2 + 1)^2};\quad
a =\frac{\hbar^2}{m\alpha}=\,\,{\rm Bohr~radius}
\end{equation}
and use both to work out the expectation value,
\begin{eqnarray}
\Delta E&=&\langle b^2R^2/2m-\alpha b/P\rangle=
 \frac{2\pi b^2}{m}\int_0^\infty (r\psi(r))^2\,dr - \frac{4\pi\alpha b}
 {h^3}\int_0^\infty p\phi^2(p)\,dp\nonumber \\
 &=& 3b^2a^2/2m - 16\alpha ba/3\pi\hbar \simeq -16b\hbar/3\pi m,
\end{eqnarray}
to first order in $b$.

With the variational method adopt a trial wavefunction just like 
eq. (16), except that $a$ is no longer identified with the Bohr radius.
Then one gets (as a function of $a$),
\begin{equation}
E(a) = \frac{\hbar^2}{2ma^2} + \frac{3b^2a^2}{2m} -\frac{\alpha}{a}
       - \frac{16\alpha ba}{3\hbar\pi}.
\end{equation}
Minimising the energy, a cubic equation for the trial radius $a$ is found:
\begin{equation}
\frac{\hbar^2}{ma^3}-\frac{\alpha}{a^2}+\frac{16\alpha b}{3\hbar\pi}=0,
\end{equation} 
whose solution to order $b$ is $a\simeq\hbar^2/m\alpha +
 16b\hbar^5/3m^3\alpha^3$. Hence the optimal energy is obtained as
$$ E =-m\alpha^2/2\hbar^2 - 16b\hbar/3\pi m + \ldots,$$
thereby agreeing with (17).

Taking the H-atom as the archetypical case we must make sure that
atomic values are hardly touched by the reciprocity term $\alpha b/P$. 
The Rydberg energy of 13.6... eV $\sim 10^{-18}$~J is known to nine 
places of decimals and because $\Delta E \sim b\hbar/m_e\sim 10^{-4}b$~J
we shall require this to be $\leq 10^{-30}$~J, just to be on the safe side.
This sets a limit $b < 10^{-26}$~kg/s. This may seem miniscule but if we
translate it into a universal angular frequency via $\Omega \equiv b/m_e$,
we obtain a value $\Omega \leq 10^4$~Hz that is not so small. As we shall 
see presently, galactic considerations produce periods which are {\em many}
orders of magnitude greater; therefore we can state with some confidence
that the reciprocity modification with small enough $b$ has essentially 
no effect on the accuracy of atomic calculations.

\section{Reciprocity and rotation of the galaxy}
We now treat the gravitational case, based on classical Newtonian dynamics 
and, presently, its reciprocity variant (3). It is a really important subject 
as the existence of possible dark matter owes a great deal to this study. 
In order to make any headway, observations of the motion/distribution of 
(visible) matter have to be taken into account \cite{Astro1,Astro2}. It has 
been established that 
the Milky Way contains a spinning supermassive black hole at centre with 
a mass of at least $4\times 10^7$ suns or about $10^{38}$~kg.  Visible mass 
is about $2\times 10^{11}$ suns and on the basis of perceived rotation rates
it is inferred that three or four times that mass is hidden in dark
matter, as far out as we can see---{\em assuming ordinary Newtonian} gravity.
The rotation speed of the Milky Way \cite{C} seems to increase linearly with
radius over a small distance, of about 0.3 kpc $\sim 10^{19}$ m, peaks,
oscillates a bit and settles down to some 230 km/s out to $5\times10^{20}$ m
from the black hole at the centre. We shall take this as a crude description
of our galaxy, neglecting spiral arm structure and the effect of some 
barring of mass in the middle on the motion.
We wish to investigate whether Born's reciprocity modification (3)
has anything of consequence to say on the topic of dark matter and, in
particular, if the observed tangential velocity profile is consistent with 
the {\em visible} matter distribution, without invoking dark matter.

Since we are presuming that the rotation is mainly tangential, if we
consider a typical star such as the sun, we may take $p_r\simeq 0$ to
a good approximation. So we just need to sum over all other masses $M$ 
and their relative speeds $v$ to obtain an average value for 
$\langle\alpha(1/r + b/p)\rangle = \langle GmM(1/r + \Omega/v)\rangle$.
Here we have adopted a positive $b$-sign, with $b\equiv m\Omega >0$, as
before, to fit in with the observations to come; the opposite sign gives 
nothing but grief. 

We cannot perform such averaging without some idea of the mass density 
distribution $\rho({\bf r})$. {\em Unmodified} Newtonian gravity indicates
that the matter distribution is roughly {\em spherical} and $\rho(r) 
\sim 1/r^2$ as far out as one can see, to ensure that the enclosed mass
increases linearly with radius and reproduce a constant tangential
star speed; hence dark matter. Of course the visible mass distribution
flatly contradicts this: it is mostly disk-shaped, with a concentration 
near the galaxy centre. This is what we shall model by taking a cylindrical
Gaussian distribution,
\begin{equation}
\rho(\varrho,z)\simeq\kappa^2\delta{\cal M}
      {\rm e}^{-\kappa^2(\varrho^2+\delta^2z^2)}/\varrho\pi^2,
\end{equation}
to which we shall add a contribution from the central bulge including a black
hole. (Here $\varrho$ is the distance from the central axis and $z$ is the 
distance from the galactic plane.) The integral over the visible mass 
density produces a finite galactic mass $\cal M$; the parameter $\kappa$ 
specifies the size of the galaxy disk while $\delta$ is the ratio of disk 
diameter to disk thickness which varies from about 10 near the centre to
50 at the outer reaches of the disk; so we will take an average value
$\delta^2\simeq 1000$ presently.

Firstly we derive the potential energy in the plane of the disk at
location $(\varrho,z=0)$ for the cylindrical mass approximation above.
\begin{eqnarray}
V(\varrho)&=&-\frac{Gm{\cal M}\kappa^2\delta}{\pi^2}
 \int\!\int\!\int d\phi dz' d\varrho' \,
 \frac{{\rm e}^{-\kappa^2(\varrho'^2+\delta^2 z'^2)}}
 {\sqrt{\varrho^2+\varrho'^2-2\varrho\varrho'\cos\phi+z'^2}}\nonumber\\
 &\simeq& -\frac{2Gm{\cal M}\kappa^2\delta}{\pi}
 \int\!\int\!dz' d\varrho' \,
 \frac{{\rm e}^{-\kappa^2(\varrho'^2+\delta^2 z'^2)}}
 {\sqrt{\varrho^2+\varrho'^2+z'^2}} \nonumber \\
 &=& -\frac{Gm{\cal M}\kappa}{\sqrt{\pi}}\int_0^\infty
 \frac{{\rm e}^{-u\kappa^2 \varrho^2}\,du}{\sqrt{u(1+u)(1+u/\delta^2)}},
\end{eqnarray}
where we have used the integral representation $2\int_0^\infty
{\rm e}^{-\xi^2 X}\,d\xi = \sqrt{\pi/X}$.
For large $\varrho$ note that $V(\varrho)\simeq -Gm{\cal M}/\varrho$ as 
expected for a finite-sized source, while for small distances
$V(\varrho) \simeq 2Gm{\cal M}\kappa^2\delta\varrho$ vanishes as $\varrho
\rightarrow 0$. To (21) we will shortly add a contribution from the 
central bulge, approximated by a mass $c{\cal M}$ placed at the middle.

The average over relative velocity $v$ is a lot cruder and relies on data
amassed over many years by astronomers. The galactic disk \cite{Astro1}
exerts a strong attraction towards the plane on stars which stray from the 
disk and this results in a velocity dispersion $\Delta v_z \sim 30$ km/s. As we 
are supposing that the majority of stars are turning at the same orbital speed 
$v$, the only relevant quantity is the angle $\phi$ between the azimuths of 
the two bodies and the velocity dispersion along the $z$-axis, 
$|{\bf v'}-{\bf v}| = \sqrt{(2v\sin(\phi/2))^2+4\Delta v_z^2}$. Therefore 
the sum simplifies to an integration over azimuth:
$$\int_0^{2\pi} \frac{d\phi}{2\pi|{\bf v}'-{\bf v}|} \simeq\frac{1}{4\pi v}
\int_0^{2\pi}\,\frac{d\phi}{\sqrt{\sin^2{\phi/2}+\epsilon^2}}
 = \frac{K(\frac{1}{1+\epsilon^2})}{\pi v\sqrt{1+\epsilon^2}}
\equiv \frac{L}{v},$$
where $K$ is the elliptic integral of the first kind and $\epsilon\equiv
\Delta v_z/v \sim 0.15$ roughly \cite{Astro1}. (We use $L$ as a 
parameter to fit the data in due course.) So without much compunction we 
shall simply take $\langle\Omega/|{\bf v}-{\bf v}'|\rangle \sim L\Omega/v$ 
as a fair approximation; after all we are only striving to get an idea of 
the size of the reciprocity constant $\Omega$ here.

Finally there is the matter of the halo contribution. Unlike mainstream
ideas inferring dark matter, we rely on observations of {\em visible} matter 
(population II stars, white or brown dwarfs, star clusters, etc.) to put a 
bound of about $c_H \simeq$ 10\% on the amount of halo mass $c_H{\cal M}$.
Furthermore we shall suppose, like everyone else, that the halo velocity 
distribution is largely thermal \cite{Astro2}.
In order to model these effects we will neglect the small oblateness of
the halo and use a radial halo density distribution,
$\rho_H(r) \simeq c_H\kappa_H{\cal M}/2\pi^2r^2(1+\kappa_H^2r^2),$
multiplied by a normalized velocity distribution $\rho_H(v) = 
(\beta/\pi)^{3/2}{\rm e}^{-\beta v^2}.$
This fixes the mass of the halo to be $c_H{\cal M}$ and the halo velocity 
dispersion \cite{Astro2} to be $(\Delta v_H)^2 = 3/2\beta$, while the size 
of the halo is determined by 1/$\kappa_H$. It allows us to work out the 
contribution to the gravitational potential energy due to
the halo:
\begin{eqnarray}
V_H(r)&=&-Gm\left[\frac{1}{r}\int_0^r\rho_H(r')\,4\pi r'^2dr' +
         \int_r^\infty \rho_H(r')\,4\pi r'dr'\right] \nonumber \\
 &=&  -\frac{Gmc_H\kappa_H{\cal M}}{\pi}\left[\frac{2}{\kappa_Hr}
    \arctan(\kappa_Hr) + \ln(1+\frac{1}{\kappa_H^2r^2})\right],
\end{eqnarray}
as well as the reciprocity halo contribution:
\begin{equation}
\langle\frac{1}{v}\rangle_H = \frac{1}{v}\int_0^v\rho_H(v')\,4\pi v'^2dv'
     +\int_v^\infty\rho_H(v')\,4\pi v'dv' = {\rm erf}(\sqrt{\beta}v)/v.
\end{equation}

Putting all this together, and including reciprocity terms, we obtain 
the total energy dependence on angular momentum $\ell = m\varrho v$ 
(conserved by axial symmetry) and distance from axis $\varrho$ for a 
test mass $m$:
$$E(\varrho,\ell)\!=\!\frac{\ell^2}{2m\varrho^2}
+\frac{1}{2}m\Omega^2\varrho^2
-\!(L+c)\frac{Gm^2{\cal M}\Omega\varrho}{\ell}-Gm{\cal M}\!\left(
\frac{c}{\varrho} + \frac{\kappa\delta}{\sqrt{\pi}}\int_0^\infty\!\!
\frac{{\rm e}^{-u\kappa^2\varrho^2}\,du}{\sqrt{u(1+u)(\delta^2+u)}}\right)$$
$$\qquad -\frac{Gm{\cal M}c_H\kappa_H}{\pi}
\left[\frac{2}{\kappa_Hr}
    \arctan(\kappa_Hr) + \ln(1+\frac{1}{\kappa_H^2r^2})\right]-
\frac{Gm^2{\cal M}c_H\Omega\varrho}{\ell}{\rm erf}(\frac{\sqrt{\beta}\ell}
 {m\varrho}).$$
Since we are presuming that the velocity is largely tangential, 
$\dot{p_\varrho} = 0 = \partial E/\partial\varrho$, which leads to the force
equation:
\begin{eqnarray}
0 &=& -\frac{mv^2}{\varrho}+m\Omega^2\varrho-(L+c)Gm{\cal M}\frac{\Omega}
{v\varrho} +\nonumber \\
&& + Gm{\cal M}\left(\frac{c}{\varrho^2}+
 \frac{2\kappa^3\varrho}{\sqrt{\pi}}
 \int_0^\infty\sqrt{\frac{u}{1+u)(1+u/\delta^2)}}
{\rm e}^{-u\kappa^2\varrho^2}\,du \right) \nonumber \\
&& +\frac{2Gm{\cal M}c_H}{\pi\varrho^2}\arctan(\kappa_H\varrho)
  \!+\!\frac{Gm{\cal M}\Omega c_H}{\varrho}\left[
  2\sqrt{\frac{\beta}{\pi}}{\rm e}^{-\beta v^2}\!\!-\!\!
    \frac{{\rm erf}(\sqrt{\beta}v)}{v}\right].
\end{eqnarray}

We can simplify the look of this equation by rescaling to 
dimensionless variables. Let ${\cal V} = v/\sqrt{G{\cal M}\kappa},\, 
{\cal R} = \kappa\varrho,\,\omega = \Omega/\sqrt{G{\cal M}\kappa^3},
{\cal B} =\beta\kappa G{\cal M}$. The tangential
velocity profile then reduces to solving an equation for ${\cal V}$ as a 
function of ${\cal R}$:
$$\hspace{-4cm}{\cal V}^2 + \frac{\omega(L+c)}{\cal V}+
 c_H\omega\left[\frac{{\rm erf}(\sqrt{\cal B}{\cal V})}{\cal V}
 - 2\sqrt{\frac{\cal B}{\pi}}\,\,{\rm e}^{-{\cal B}{\cal V}^2}\right] $$
\begin{equation}
=\omega^2{\cal R}^2
+\left[\frac{c}{\cal R}+\frac{2{\cal R}^2}{\sqrt{\pi}}\int_0^\infty\!\!
\frac{\sqrt{u}\,{\rm e}^{-u{\cal R}^2}\,du}{\sqrt{(1+u)(1+u/\delta^2)}}
\right] +\! \frac{2c_H}{\pi{\cal R}}\arctan
\left(\!\frac{\kappa_H{\cal R}}{\kappa}\!\right).
\end{equation}
The chosen sign for $\Omega$ is highly significant in determining the 
behaviour of the velocity for small $\cal R$, viz. 
${\cal V}\rightarrow (L+c){\cal R}\omega/c$, 
as required by the data. [Had we reversed the sign of $\omega$ or $b$ we 
would not have been able to fit observations even remotely.]

A numerical solution of (23) is possible once a few measured values
are input. The galaxy is 50000 light-years or more in radius, i.e. about
$5\times 10^{20}$~m, so let us set the scale 
$\kappa \sim 0.7\times10^{-20}$ m$^{-1}$ 
or so. Taking the visible galactic mass to be roughly ${\cal M}\sim 2\times 
10^{41}$~kg, one estimates that $\kappa^3G{\cal M}\sim 5\times10^{-30}$/s$^2$.
Also the rotation speed outside the innermost part of the galaxy
equals about 230 km/s and $v$ grows to this value over about 0.3 kpc or 
$10^{19}$ m, telling us that $\Omega(L+c)/c\sim 2.5\times 10^{-14}$
s$^{-1}$. In attempting to fit the observed velocity profile, use values
$c\simeq 0.05$, as the proportion of galactic mass concentrated in the
central bulge, and $c_H =0.01$ as the ratio of halo mass to disk mass.
For simplicity take the halo radius to equal the disk radius or $\kappa
=\kappa_H$; although this is an underestimate it makes little difference 
to the numerical results because $c_H$ is quite small anyhow. A similar 
comment attaches to our chosen $\cal B$-value of about 13 \cite{Astro2}.
Finally set $L\simeq 0.85$ and $\Omega \simeq 1.5\times 10^{-15}$ Hz,
implying $\omega \simeq 0.2$. None of these inputs is at all absurd.

The profile equation (25) now determines ${\cal V}$ as a function of 
${\cal R}$ and the numerical results are plotted in Figure 4. 
In trying to fit the data it is important to mention that there are
two roots for the cubic in $\cal V$. In the inner region we use the 
smaller root, corresponding to the linear relation $v \simeq 
r\Omega (L+c)/c$, and beyond about $r = 10^{19}$ m we adopt the larger
root. Although the Figure 4 graph is rather high below 2 kpc and shows a 
drop off to steadyish speed 
that is a bit faster than what is observed, the general shape of the 
curve is moderately satisfactory. Moreover the flattish part of the curve 
has $v\sim 200$ km/s, which is the correct magnitude. However it must be
admitted that our model and calculation have many rough edges and need
a lot more refinement before they can be judged a success; it even
conceivable that dressing the model may spoil it rather than enhance it.
\begin{figure}
\begin{center}
\epsfbox{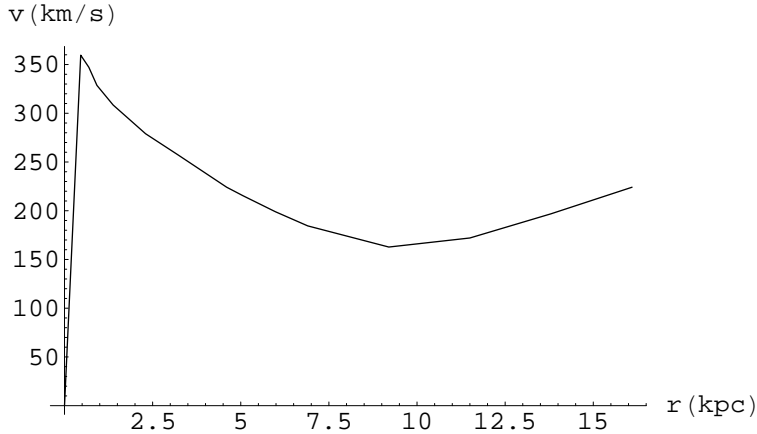}
\end{center}
\caption{\label{Fig4Label}Predicted tangential velocity curve with radius.}
\end{figure}

\section{Conclusions and criticisms}
The above fit has some good features and some bad ones; on the
positive side it is able in principle to produce an orbital velocity
curve that matches the data near the centre (linear rise) and at large 
distances (almost constant speed), without calling upon a dark matter 
component --- Born's reciprocity principle has after all modified 
Newtonian dynamics in a radical way. But it must be confessed that 
the detailed fit above is far from convincing.
This negative feature may be due to the crudeness of our visible mass
and velocity distributions. It would be more realistic to include extra
matter in the inner galactic bulge, introduce barring plus spiral arms and
generally model the velocity distribution more accurately when working
out the gravitational potential and its reciprocity counterpart.
This is clearly fertile ground for future research.
Meanwhile we can probably be content with our {\em estimate} of 
$\Omega \simeq 1.5\times 10^{-15}$ Hz for Born's hypothesis and not 
much more; it is truly a galactic scale and corresponds to a period of 
1.3$\times 10^{8}$ years, which may be connected with the rotation rate of
the galaxy.

Serious concerns remain. The scheme destroys translational 
invariance because of the way that momentum and position are tied, so
a violation of Galilean relativity is to be expected, never mind
Lorentzian relativity.  Choice of origin is another issue. We have picked 
the galactic centre as the obvious place; however one might fret 
about effects of harmonic acceleration ($\Omega^2 {\bf r}$) which can get 
overwhelmingly large from outer regions of the universe, but is somewhat
insignificant within a galaxy. Fortunately it appears that on cosmic 
scales the galaxies are uniformly distributed in every direction
all around so one may presume that such attractions will cancel out 
overall. And because galaxies are receding away from us at Hubble rates 
the relative velocity term $\Omega/v$ diminishes and balances out as one
goes out. So it would seem, superficially at least, that one could choose any 
other galaxy and take the origin at its centre to reproduce its own galactic
rotation without worrying unduly about other galaxies.

Fundamental theoretical criticisms can nevertheless be levelled at (3).
It is explicitly non-relativistic and should at least be made to conform with 
special relativity, To carry out that program sensibly one would 
need to study the quaplectic group \cite{L} and augment the electromagnetic
or gravitational field (photon/graviton exchange) by their reciprocity 
analogues or some other contributions. This signifies overhauling the whole of 
standard field theory and the task seems rather difficult, if not 
vague, at this stage. In the end it may turn out that Born reciprocity
is unable to fit the data properly or mesh in with our familiar relativistic
field theory concepts.  Even if it works in limited fashion it would have to 
be seen as one of a panoply of Modified Newtonian Descriptions (MOND) of 
gravity over large distances. Taking a skeptical point of view, 
Born's idea will probably be found wanting, dark matter will be required and 
the problem of seeing it by non-gravitational methods will remain with us 
for a good while. But before the death knell is finally sounded on the subject
of reciprocity, specialist galactic modellers need to investigate its
ramifications comprehensively.

\section*{Acknowledgements}
We would like to thank Professor John Dickey for enlightening discussions on 
galactic structure and for pointing us in the right direction.

\end{document}